\documentclass[aps,prl]{revtex4}

\usepackage{amssymb}
\usepackage{amsmath}
\usepackage{graphicx}
\usepackage{float}

\begin{document}

\title{Complexity and transition to chaos in coupled Adler-type oscillators.}

\author{D. \surname{Estevez-Moya}}
\affiliation{Facultad de F\'isica, Universidad de La Habana, San Lazaro y L. CP 10400. La Habana. Cuba.}

\author{E. \surname{Estevez-Rams}}
\email{estevez@fisica.uh.cu}
\affiliation{Facultad de F\'isica-Instituto de Ciencias y Tecnolog\'ia de Materiales(IMRE), Universidad de La Habana, San Lazaro y L. CP 10400. La Habana. Cuba.}

\author{H. \surname{Kantz}}
\affiliation{MPI for the Physics of Complex Systems. N\"othnitzer Strasse
  38. D-01187 Dresden.}

\begin{abstract}
  Coupled non-linear oscillators are ubiquitous in dynamical studies. A wealth of behaviors have been found mostly for globally coupled systems.  From a complexity perspective, less studied have been systems with local coupling, which is the subject of this contribution. The phase approximation is used, as weak coupling is assumed. In particular, the so called needle region, in parameter space, for Adler-type oscillators with nearest neighbors coupling is carefully characterized. The reason for this emphasis is that in the border of this region to the surrounding chaotic one, computation enhancement at the edge of chaos has been reported. The present study shows that different behaviors within the needle region can be found, and a smooth change of dynamics could be identified. Entropic measures further emphasize the region's heterogeneous nature with interesting features, as seen in the spatiotemporal diagrams. The occurrence of wave-like patterns in the spatiotemporal diagrams points to non-trivial correlations in both dimensions. The wave patterns change as the control parameters change without exiting the needle region. Spatial correlation is only achieved locally at the onset of chaos, with different clusters of oscillators behaving coherently while disordered boundaries appear between them.
\end{abstract}


\date{\today}
\maketitle

\begin{quotation}

\end{quotation}

\section{Introduction}

Coupled non-linear oscillators under weak coupling conditions have been studied, at least since the pioneering works of Winfree \cite{winfre67}. In his approach, the oscillator phase is the relevant parameter in the dynamics instead of the oscillator amplitude. Earlier studies by Adler \cite{adler46} of locking occurrence in feedback circuits lead to an equation for a type of non-linear oscillators (since then known as Adler-type oscillators) with phase equation
\begin{equation}
 \frac{d\theta}{dt}=-\gamma \sin \theta +\Delta \omega, \label{eq:adler}
\end{equation}
where $\theta$ is the oscillator phase, $\gamma$ a control parameter, and $\Delta \omega=\omega_0-\omega_1$ is the difference between the natural frequency of the oscillator and the imposed frequency on the circuit. Adler's work has been generalized to systems of coupled non-linear oscillators that describe a wide range of phenomena in diverse fields ranging from physics to chemistry, biology, and neurobiology, among others \cite{winfre67,kuramoto75,strogatz01,mosekilde02,zillmer06}. When the system of non-linear oscillators has a weak global coupling between the oscillators described by a harmonic signal, it is known as the Kuramoto model \cite{kuramoto75}. This type of system has been used to study synchronization \cite{acebron05}.

Recently, a system of locally coupled non-linear Adler-type oscillators has been proposed as a toy model that exhibits different chaotic, regular, and complex behaviors depending on the control parameters \cite{alonso17}. The local coupling is realized, taking the oscillators as a ring with nearest-neighbor interaction described by an interaction parameter. The phase diagram of the system, depending on two control parameters, has been reported \cite{alonso17,estevez18}. The most striking fact of the phase diagram is the emergence of a rich set of behaviors that reminds of those found in cellular automata. In the phase diagram, several distinctive regions have been identified \cite{estevez18}:
\begin{itemize}
 \item One region exhibits near zero entropy which has been called the absorbing region;  
\item a second region has chaotic behavior showing disordered states and high sensitivity to initial conditions with no spatiotemporal correlations; 
\item a region termed wedge shows entropy values above zero but not as high as the chaotic region with spatiotemporal diagrams exhibiting long-range spatial and temporal correlations; 
\item a fourth region, termed the needle region, is identified with intermediate values of disorder, sensibility to initial conditions, and striking spatiotemporal patterns. 
\end{itemize}

Additional to this rich set of behaviors, further work on the system has unveiled a transition to a chaotic regime, where, it has been reported, the system of oscillators allows an enhancement of computation at the edge of chaos \cite{estevez19}. Dynamical systems can be viewed as computational devices capable of storage, transmission, and production of information \cite{crutchfield12}. The hypothesis of computation enhancement near the edge of chaos (EOC)  states that, in the transition towards a chaotic regime, some dynamical systems exhibit augmented capabilities as a computational device \cite{su89,sole96,bertschinger04,beggs08,boedecker12}.  

In this contribution, we study the needle region, where EOC has been found, while the system moves towards the chaotic region. The paper is organized as follows. The mathematical model is formally introduced in section \ref{sec:model}, and some previous results are summarized. Section \ref{sec:methods} introduces the entropic measures used to characterize the system behavior. Section \ref{sec:results} presents the main results, followed by a section \ref{sec:discus} of discussion and, finally, the conclusions.

\section{Nearest neighbor coupled Adler-type oscillators}\label{sec:model}

A simple model for ensembles of active elements is to consider $N$ oscillators of Adler-type, arranged in a ring coupled with the nearest neighbors oscillators and governed by a system of equations given by,
\begin{equation}
 \frac{d\theta_i}{dt}=\omega+\gamma \cos \theta_i+(-1)^ i k \left [ \cos \theta_{i-1}+\cos \theta_{i+1}\right ].\label{eq:mastereq}
\end{equation}
$\theta_i$ is the phase of each oscillator, which varies with time; $\omega$ describes the natural frequency of the oscillators when no feedback and coupling are present. The control parameters are $\gamma$, which determines the strength of the self-feedback, and $k$, giving the coupling strength. The alternating sign for the coupling term is meant to balance the feedback avoiding drifts in the oscillator behaviors due to a bias in the chain. All control parameters are taken non-negative without loss of generality. The observable magnitude is considered to be $\sin \theta_i$, called the activity, \footnote{$\cos \phi$ could be used instead, which is, given eq. (\ref{eq:mastereq}), a more ''natural'' choice. Instead, we decided to follow the convention of previous works \cite{alonso17,estevez18}.};

In vector form, equation ({\ref{eq:mastereq}}) can be written as
\begin{equation}
\mathbf{\dot{\theta}}=\omega \mathbf{I_N}+\gamma \mathbf{cos[ \theta]}+ k \mathbf{C}\cdot \mathbf{cos[ \theta]}.\label{eq:vectadler}
\end{equation}
where $\mathbf{\dot{\theta}}$, is the vector of $\dot{\theta}_i$, $\mathbf{I_N}$, is the $N$ components identity vector, $\mathbf{cos[\theta]}$ is a $N$ vector with components $\cos \theta_i$, $\mathbf{C}$ is a $N\times N$,  with zero main diagonal, skew-symmetric, tridiagonal matrix with alternating $\pm 1$ along the super- and sub-diagonals. 

For $k=0$, no coupling exists. The system behaves as independent, pure Adler oscillators.  
\begin{equation}
 \frac{d\theta}{dt}=\omega+\gamma \cos \theta,\label{eq:adler1}
\end{equation}
where a positive value of $\cos \theta$ implies a faster phase velocity. The opposite, slowing of the phase velocity, occurs when $\cos \theta<0$ and the right-hand side of equation (\ref{eq:adler1}) is still positive. In such a case, the oscillator is going through a bottleneck. If $\gamma \geq \omega$,  when $\cos \theta=-\omega/\gamma$ the phase is locked, $d\theta/dt=0$. Phase locking will happen for two values of $\theta$; one is stable, and the other unstable. When $\gamma<\omega$, no phase locking can happen, but the system, behaving periodically, still shows intervals of $\theta$ where bottleneck slowing of the phase velocity happens. In this interval, the system will spend most of its time within a period.

For $k\neq 0$, the coupling term relates the phase speed of an oscillator with its neighbors. The phase velocity of oscillator $i$ increases if
the self-feedback and the coupling terms are positive due to the strengthening contribution of neighbors. If the two terms are negative, but the right-hand side of the equation (\ref{eq:mastereq}) is still positive, the oscillator $i$ is in a bottleneck. If the right-hand side is negative, the phase velocity reverses sign, and its absolute value increases with an increasing absolute value of the self-feedback and coupling term. If $\cos \theta_{i-1}$ and $\cos \theta_{i+1}$ have different signs, they have competing effects over the phase velocity. 

The coupling has alternating signs, being positive (negative) for even (odd) numbered oscillators, so taking $N$ as an even number guarantees balance in the sign of the interaction for the whole system. The reader is referred to \cite{alonso17,estevez18} for further details.

The analysis of the existence and stability equilibrium points has been thoroughly made by Alonso \cite{alonso17}. A summary follows. Making $\mathbf{\dot{\theta}}=\mathbf{0}$, and solving the linear system of equations (\ref{eq:vectadler}),  the equilibrium points are given by
\begin{equation}
 \begin{array}{l}
\cos \theta_{2i+1}^*=a=-\omega \frac{\gamma+2k}{\gamma^2+4k^2},\\\\  
\cos \theta_{2i}^*=b=-\omega\frac{\gamma-2 k}{\gamma^2+4 k^2},\\\\\label{eq:equilibrium}
 \end{array}
\end{equation}
from which equilibria exist if and only if,
\begin{equation}
 \omega \leq \frac{\gamma^2+k^2}{\gamma+2 k}.\label{eq:eqregion}
\end{equation}
From there, when $|a|<1$, there are $2^N$ possible equilibrium points, equally split between odd and even oscillators. For $|a|=1$, $\theta_{2i-1}=\pm\pi$, the odd components are identical, and there are $2^{N/2}$ equilibria. The bifurcation curve is given by the equality in (\ref{eq:eqregion}).

Each of the equilibria undergoes a saddle-node bifurcation yielding a pair of equilibria, with the smooth field induced by equation (\ref{eq:vectadler}) restrained to the direction that joins each pair of equilibria. 

Stability has been studied numerically, and only one point was found to be stable. It corresponds to all values of $\theta_i$ positive. The probability of another point being stable was smaller than $10^{-6}$. The behavior of the largest eigenvalue of the corresponding family of Jacobians has been discussed in \cite{alonso17}, and we refer the reader there for further details. The conclusion is that the presence of equilibria alone, as discussed, may not be sufficient to explain the observed complex behavior within the region with no fixed points. Therefore, the occurring bottlenecks near bifurcation may also play a role in the overall dynamics.

Chaotic regimes can be in time or the spatial coordinate (oscillators index) \cite{chian13}; this has been studied by Estevez et al. \cite{estevez18} for the system given by equation (\ref{eq:vectadler}). Low values of spatial entropy density can only accommodate low values of temporal entropy density. In the needle region (see figure \ref{fig:clz}), temporal entropy density is near zero. A plot of temporal entropy density vs. spatial entropy density (See figure 4 in \cite{estevez18}) shows the distribution of points below the diagonal of equality between both measures, temporal and spatial, of disorder.

The system can be rescaled in terms of $k$ by changing the time units. From now on, it will be taken $k=1$ in equation (\ref{eq:mastereq})  without loss of generality. The parameters $(\omega, \gamma)$ will be taken as real numbers varying continuously. A schematic of the parameter space diagram (drawn as a function of $\omega$ and $\gamma$) shows different regions (Figure \ref{fig:clz}), which will be explained after we introduce the entropic magnitudes. 

\section{Data processing and entropic magnitudes}\label{sec:methods}

In the results that follow, the system of differential equations given by (\ref{eq:mastereq}) is solved numerically using a Runge-Kutta method of order $4$, as implemented in the GNU Scientific Library \cite{gsl}. In order to discover any discrepancy in the stability of the numerical solutions, comparisons were made, for several instances of the problem, with the solution found by using Wolfram Mathematica \cite{Mathematica} numerical differential equation solver. No discrepancy was found. Simulations were performed over $N=5 \times 10^3$ oscillators. To test the dependence of the result on the system size, for specific values of $(\omega, \gamma)$, $N$ was taken as $2\times 10^5$, observing no change in the calculated values. Averaging over $100$ runs of different random initial conditions was performed.

The system is left to evolve for $t=N$ steps. The short-range coupling imposes at least $N/2$  times steps to reach the rest of the chain from a signal originating at any oscillator. $N$ proved to be a suitable number of steps to kill all possible transient behavior. Activity $\sin \theta_i$ values were binarized, taking the mean value as a threshold. In such a way, the binarization procedure does not guarantee the same number of $1$s or $0$s. After binarization, the spatial configuration of the oscillators will be described by a string $s$ of length $N$. 

Shannon entropy rate $h$ will be used to measure information production. Consider a bi-infinite sequence $S^{(t)}=\ldots s^{(t)}_{-2}s^{(t)}_{-1}s^{(t)}_{0}s^{(t)}_{1}s^{(t)}_{2}\ldots$, where $s^{(t)}_i$ is the binary symbol observed in the cell $i$ at time step $t$. The entropy density can be considered the amount of new information in the observation of cell $s_i$, conditional on the state of all previous cells $s_j$, $j<i$ \cite{crutchfield03}: 
\begin{equation}
 h=H(s_i|\ldots, s_{i-1}),
\end{equation}
where $H(X|Y)$ denotes the Shannon conditional entropy of random variable $X$ given variable $Y$ \cite{cover06}. 

Entropy density can be written in terms of the block entropy,
\begin{equation}
 h=\lim\limits_{L\rightarrow \infty}\frac{H_S(L)}{L}.\label{eq:hmu}
\end{equation}
$H_S(L)$ is the the block entropy of length $L$ of the bi-infinite string $S$ given by
\begin{equation}
H_{S}(L)=-\sum\limits_{S^L}p(S^L)\log p(S^L).
\end{equation}
The sum goes over all binary sequences $S^L$ of length $L$, and $p(S^L)$ is the probability of observing one particular string $S^L$ in the bi-infinite string $S$. 

For, necessarily, finite data the entropy rate has to be estimated \cite{schurmann99,lesne09}. We will be using the Lempel-Ziv factorization \cite{lz76} procedure to estimate the entropy density. The details of the implementation can be found elsewhere \cite{estevez15}.

Effective complexity measure $E$ \cite{grassberger86}, also known as excess entropy \cite{crutchfield03}, measures the correlation at different scales in a process and is related to the intrinsic memory of a system. For a bi-infinite string, the effective measure complexity measures the mutual information between two infinite halves of the sequence \cite{grassberger86,crutchfield03},
\begin{equation}
\begin{array}{ll}
 E(S) &=I[\ldots, s_{-1}: s_0, s_1, \ldots],\label{eq:excess}
 \end{array}
\end{equation}
 $I[X:Y]=H[X]+H[Y]-H[X,Y]$ is the mutual information between $X$ and $Y$ \cite{cover06} and is a measure of the amount of information one variable carries regarding the other. $E$ is related to pattern production and context preservation and has been interpreted as the intrinsic redundancy of the source or apparent memory \cite{crutchfield03}. Effective measure complexity is estimated through Lempel-Ziv factorization using a random shuffle procedure \cite{melchert15}, as explained in \cite{estevez19}.

Information distance  $d(s,p)$ comes from algorithmic randomness \cite{kolmogorov65}. Consider the shortest algorithm $s^*$ capable of producing the string $s$; the length $K(s)=|s^*|$ of this program is called the Kolmogorov randomness of the string. Accordingly, $K(s|p^{*})$, known as Kolmogorov conditional randomness, is the length of the shortest program able to reproduce $s$ if the program $p^*$, reproducing the string $p$, is given. The information distance is defined by \cite{li04}

\begin{equation}
 d(s,p)=\frac{max\{K(s|p^{*}), K(p|s^{*})\}}{max\{K(s), K(p)\}}.\label{eq:dnid}
\end{equation}

The distance $d(s,p)$ measures how "hard" it is to reproduce one of the strings if the algorithm for producing the other string is known: two sequences that can be derived one from the other by a small-sized algorithm, will result in a small $d(s,p)$. We estimate $d(s,p)$ as explained in \cite{estevez15}, as
\begin{equation}
 d(s,p)=\frac{h(sp)-min\{h(s), h(p)\}}{max\{h(s), h(p)\}}.\label{eq:dlz}
\end{equation}
Two initial conditions, which are identical except for one randomly selected site $s_i$, are chosen. Each initial condition is left to evolve, and after a sufficient number of time steps, the information distance between both systems is calculated using (\ref{eq:dlz}). In this way, $d$ measures the system's sensitivity to a minimum perturbation of the initial state. The results are the average of performing such calculations for $20$ different pairs of initial conditions.  

Finally, following the probability $\rho$ of one symbol (e.g., $1's$) in the string, as it evolves in time, will be useful to account for the production (erasure) processes in the dynamics. 

To calculate all the described magnitudes, the last steps were used to avoid the influence of the transient region, and the mean value reported is taken for a number of different initial conditions. The standard deviation shows no significant spread of values.

\section{Analysis of the needle region.}\label{sec:results}

\begin{figure}[!t]
\centering
\includegraphics*[scale=0.7]{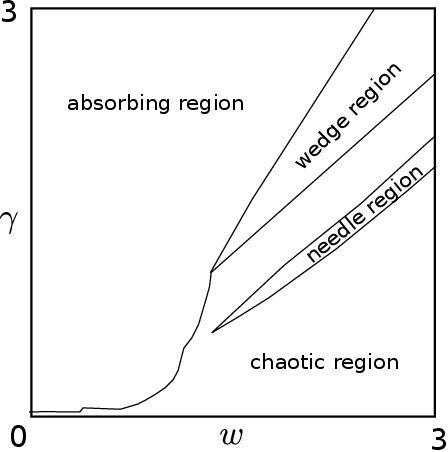}
\caption{A schematic diagram of phases in parameter space (modified from Ref. \cite{estevez18}) Four regions are identified by the entropic magnitudes. The absorbing region has zero entropy density, while the chaotic region has a value near $1$. For the wedge region, the entropy density is higher than the absorbing region but still lower than $1/2$. Low values of $h$ characterize the needle region (see Table \ref{tbl:pd}). The diagram is for the system behavior after all transient behavior has died out, as explained in  Ref. \cite{estevez18}.
}\label{fig:clz}
\end{figure}

\begin{table}[!ht]
 \center
 \renewcommand{\arraystretch}{1.5}
{
 \begin{tabular}{llllll}
\hline
region    & $h$   && $E$        && $d$ \\
\hline
absorbing & $\sim 0$    && $\sim 2.4$ && $0$\\
wedge     & $\sim 0.14$ && $\sim 6$   && $\sim 0$\\
chaotic   & $\sim 1$    && $\sim 0$   && $\sim 0.9$\\
needle    & $\sim 0$    && $\sim 4$   && $\sim 0$ \\
\hline
 \end{tabular}
}
\caption{Entropic measures in the different regions of the parameter space. $h$ stands for the entropy density, while $E$ measures the effective complexity measure. The results are the average over the last step of 100 runs of different random initial conditions. $d$ is the information distance given by equation (\ref{eq:dnid}) between a system evolving from an initial condition and the perturbed system with the same initial condition with only one oscillator phase changed. The results are the average of performing such calculations for the last steps for $20$ different pairs of initial conditions.}\label{tbl:pd}
\end{table}

Several distinct regions exist in the coupled oscillators' control parameter space (Figure \ref{fig:clz}). Table \ref{tbl:pd} summarizes the characteristics of each region (Detailed discussion can be found in \cite{estevez18}). The focus will be on the needle region, with low $h$, surrounded by the connected chaotic region with high $h$. 

\begin{figure}[!t]
\centering
\includegraphics*[scale=0.8]{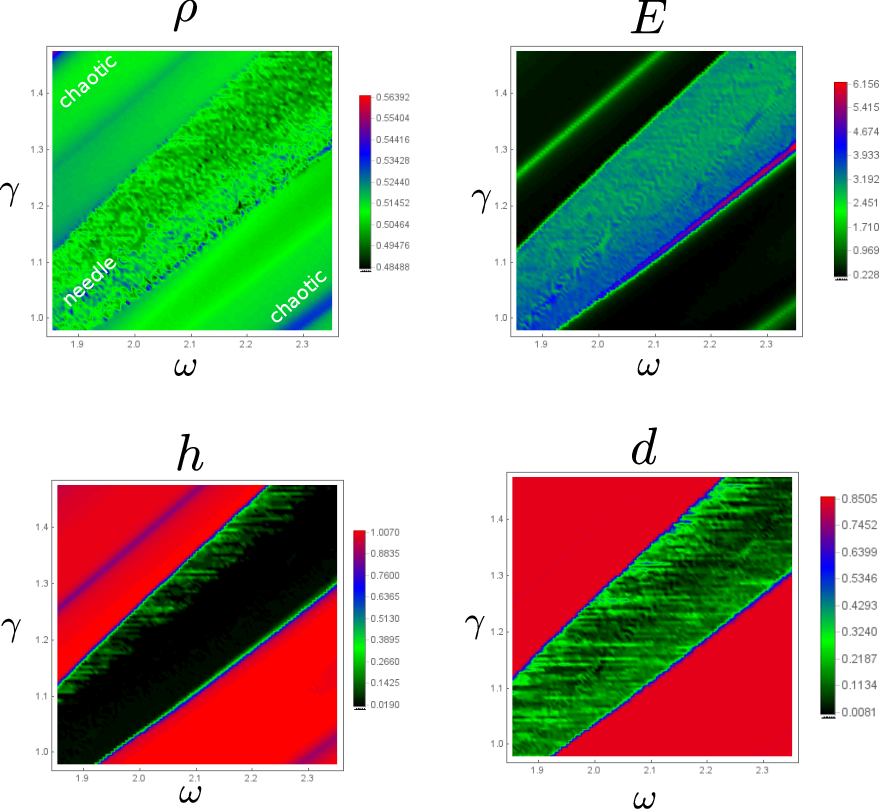}
\caption{Entropic magnitudes over a fraction of the needle region as a function of the control parameters $(\omega, \gamma)$. Calculations were performed with the last configurations, for $100$ different initial conditions, to avoid any transient behavior influence at each point, and average values are reported. The symbol density $ \rho $ shows small variations between the needle region and its chaotic surroundings, around $1/2$. Entropy density $h$ is almost zero within the needle region and near the lower boundary to the chaotic region. Correspondingly, effective measure complexity $E$ exhibits higher but inhomogeneous values at the needle region. A red strip at the lower boundary evidence a surge in $E$ at the lower boundary. The information distance $d$ is maximum in the chaotic region but above zero within the needle where a stripe pattern can be seen.   
}\label{fig:entropicpd}
\end{figure}

The are several ways to determine the exact location of the boundary to the chaotic region in a dynamical system \cite{skufca06,chian13}. In Estevez et al. \cite{estevez19}, the entropic magnitudes described in the previous section were used. At the boundary, $E$ drops to zero, while $h$ has a step increase.  Sensibility to perturbations, given by $d$, also has a jump in values. That any of these magnitudes can identify a well-localized boundary can be verified in figure \ref{fig:entropicpd} by the sharp contrast between regions.

Figure \ref{fig:entropicpd} are maps of a small region of the control space using the different entropic magnitudes. There are no significant changes in the symbol probability ($\sim 0.5$) between the needle and the chaotic region, implying that any change in the entropic variables does not come at the expense of symbol erasure. The value of $\rho\sim 1/2$ also points to the mean being near the median and hence, the distribution of $\sin \theta $ close to symmetric. Entropy density is high in the chaotic region while near zero, from almost side to side of the needle region except near the left border. At the left border, a transition area is found with entropy density above zero but well below the chaotic regime values. $d$ shows that, as expected, the chaotic behavior is characterized by high sensitivity to the initial condition. However, in the needle region, the information distance is not zero and fluctuates from zero to a value of $0.4$ in a stripe pattern with constant $\gamma$ values. The effective measure complexity further emphasizes that the needle region is far from homogeneous; wavy patterns can be seen. At the right boundary of the needle region, $E$ has a maximum (shown as a narrow red area) parallel to the boundary before it falls abruptly as it enters the chaotic regime.

\begin{figure}[!t]
\centering
\includegraphics*[scale=1]{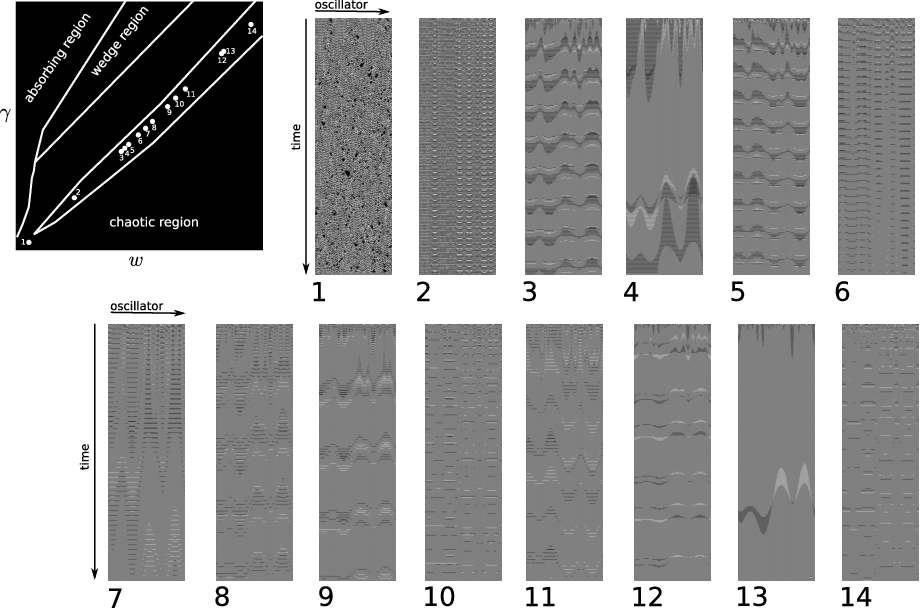}
\caption{Spatio-temporal maps along a center line in the needle region. The left upper diagram shows the points from where the spatiotemporal maps were calculated. In the spatiotemporal maps, time is on the vertical scale, and the oscillator's values are displayed horizontally in consecutive order. Point $1$ is in the chaotic region. All other points are within the needle region. The spatiotemporal maps show that, in the needle region, global correlations of the oscillators are attained.
}\label{fig:needle}
\end{figure}

To better understand the behavior in the needle region, spatiotemporal maps, starting with a random initial condition, at different points were computed along a line and are shown in figure \ref{fig:needle}. Point $1$ is still in the chaotic regime. No long-range, spatial, or temporal patterns are present in the spatiotemporal map. Points 2-14 are all within the needle region. In all cases, the initial random condition does not survive, and patterns emerge after a few time steps for the spatial arrangement of oscillators. The spatial organization does not come at the expense of the erasure of symbols as already described.

Furthermore, two correlated phenomena emerge, one in the spatial arrangement and the other in the temporal evolution. Spatial patterns are produced and evolve for a certain amount of time and then seem to disappear, and the behavior repeats itself in time. There is a characteristic wave-like image of the spatiotemporal maps. The difference between the spatiotemporal maps at different points is mainly the spatial wavelength and frequency of the pattern production and evolution. Both are correlated; the shorter the spatial wavelength, the higher the frequency. As the points are taken for increasing $\omega$ and $\gamma$ values, another cyclic behavior is observed when comparing the spatiotemporal map: spatial wavelength increases and time-frequency decreases and, at a certain point, spatial wavelength starts to decrease while frequency increases. This behavior is not periodic. 

\begin{figure}[!t]
\centering
\includegraphics*[scale=1]{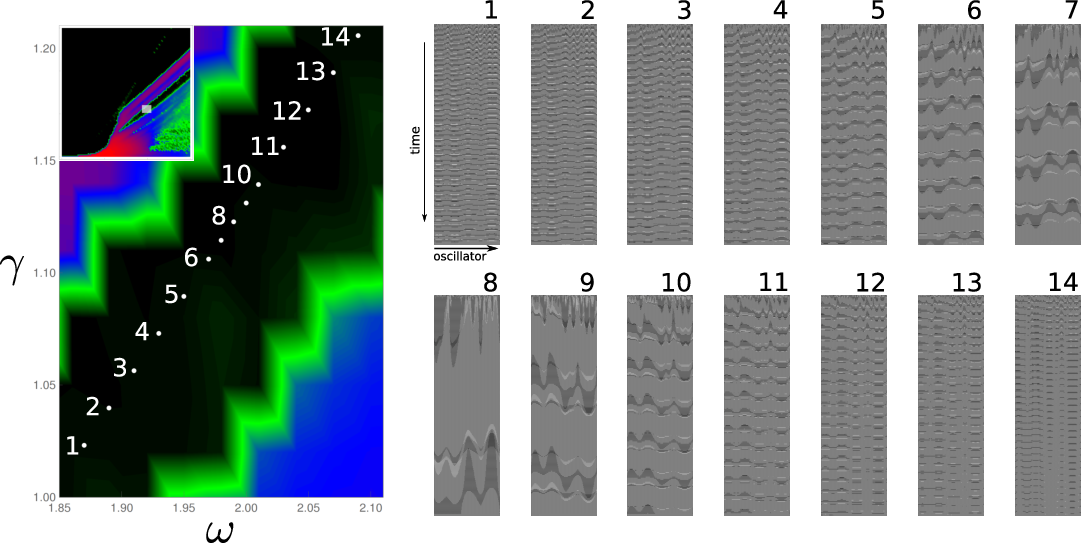}
\caption{Closer look into the spatiotemporal maps along a center line within the needle region. The left plot shows the points from where the spatiotemporal maps were calculated. Right: the spatiotemporal maps. Time is on the vertical scale, and the oscillator's values are displayed horizontally in consecutive order.
}\label{fig:needle1}
\end{figure}

One cycle is shown in figure \ref{fig:needle1}, where the evolution of the wavelength and frequency in the spatiotemporal patterns, as the system moves with increasing $ \omega $ and $ \gamma $ values, can be seen. The wavy nature of the spatiotemporal maps is clear. One must not get caught in the wrong interpretation and must consider that time increases as the configurations move downwards in the vertical axis. 

Consider any of the spatiotemporal maps. For clarity, consider as an example point 8 in figure \ref{fig:needle1} ($\omega=1.990$, $\gamma=1.125$); there is a regular background, and over the background, perturbations seem to appear at certain moments in time, at certain individual oscillators. From there, the perturbation travels spatially, perturbing their neighborhood and eventually shrinking and disappearing. The collision of two perturbations happens at the point of the extinction of the perturbation, where the oscillator returns to the background behavior. These wavy patterns indicate the spatial and temporal transfer of information among the oscillators within a time window. The difference between the points in figure \ref{fig:needle1} is the frequency with which this local structure emerges and how long they evolve before disappearing. As control parameters move from point 1 to point 8, the wavelength increases, and from there (point 9 to 14) starts to decrease. In the temporal axis, the increase in spatial wavelength is correlated with fewer initial perturbation events, but they live longer. In no case does a single local pattern last long enough to overtake the whole configuration of oscillators. 

\begin{figure}[!t]
\centering
\includegraphics*[scale=0.8]{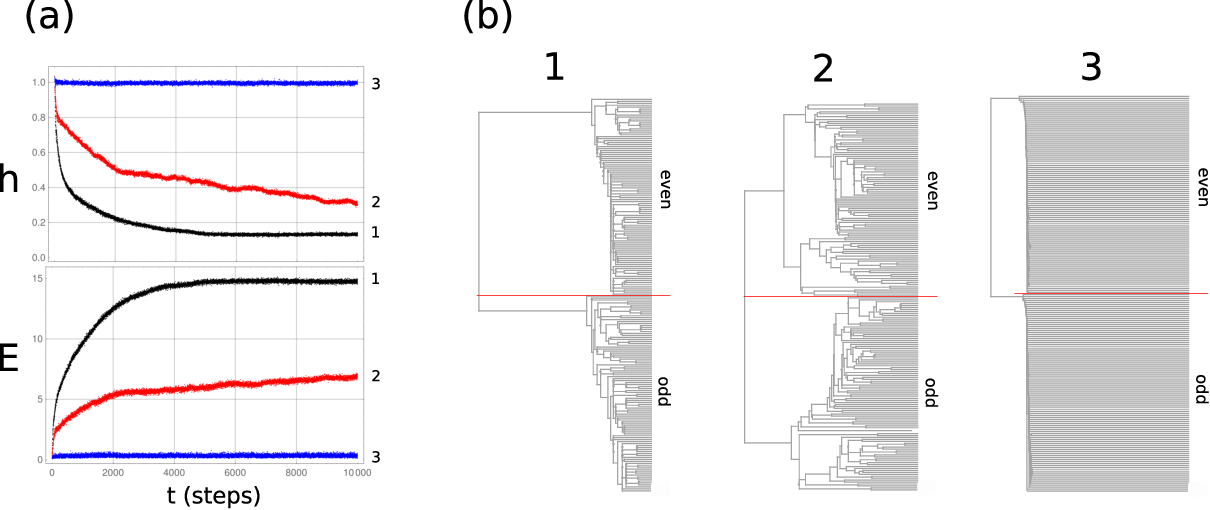}
\caption{(a) Entropy density $h$ and effective measure complexity $E$ for three points (average of $100$ different initial conditions), (1) $(\omega,\gamma)=(1.990, 1.125)$ (black) within the needle region; (2) $(\omega,\gamma)=(2.225, 1.206)$ (red) at the lower boundary of the needle region (onset of chaos) and (3) $(\omega,\gamma)=(2.500, 1.206)$ (blue) in the chaotic region. (b) Dendrogram from the distance matrix at the same points as in (a). The distance matrix $d$ is computed from the time series between every two oscillators in the system. $d$ was estimated after dropping the first $3000$ steps to cancel the influence of the transient region.
}\label{fig:plot}
\end{figure}

As shown in Figure \ref{fig:plot}a, for the chaotic point, the entropy density is constant, while it falls for the two other points. In the needle region point (curve 1 in black in figure \ref{fig:plot}a), after 6000-time steps, the system seems to have gone through the transient behavior and settled into a stationary regime, where both $h$ and $E$ do not show further variation. In contrast, the point at the boundary (curve 2 in red in figure \ref{fig:plot}a) does not show constant values of $h$ or $E$ even at $10000$ time steps. For the boundary region, at $10000$ time steps, the value of $h$ is around $0.30$, $2.14$ times the value for the needle region. The effective measure complexity increases to around $7.12$, which is more or less half the value for the needle region.

The distance matrix between the oscillators was calculated for the same three points. The matrix consists of the informational distance $d$ from the time series between every two oscillators in the system. $d$ was estimated after dropping the first $3000$ steps to cancel the influence of the transient region. The corresponding dendrogram was computed from the distance matrix and shown in figure \ref{fig:plot}b. The dendrogram tree has two branches for all points, one for the even and one for the odd oscillators. In the chaotic region, point 3, the dendrogram tree is shallow, showing that each oscillator behaves mostly independently of the others with large similarity distances ($\sim 0.8$) between them. Well within the needle region, point 1, the distance relational tree is much more complicated, with different levels and branches. Local clusters of oscillators with similar behavior, as seen by $d$, can be identified. Within a class of oscillators, odd or even, the distance between oscillators is never larger than $0.35$ showing a nearly global collective behavior of the oscillators, corroborating the visual picture given by figure \ref{fig:needle1}. The dendrogram of the system at the onset of chaos, point 2, shows a similar multilevel tree with a complicated hierarchy of branches; local clusters can also be identified, yet, compared with the needle interior, a larger distance between clusters that can go up to $0.67$, can bee seen between some clusters, showing that global coupling is not achieved. 

\begin{figure}[!t]
\centering
\includegraphics*[scale=0.8]{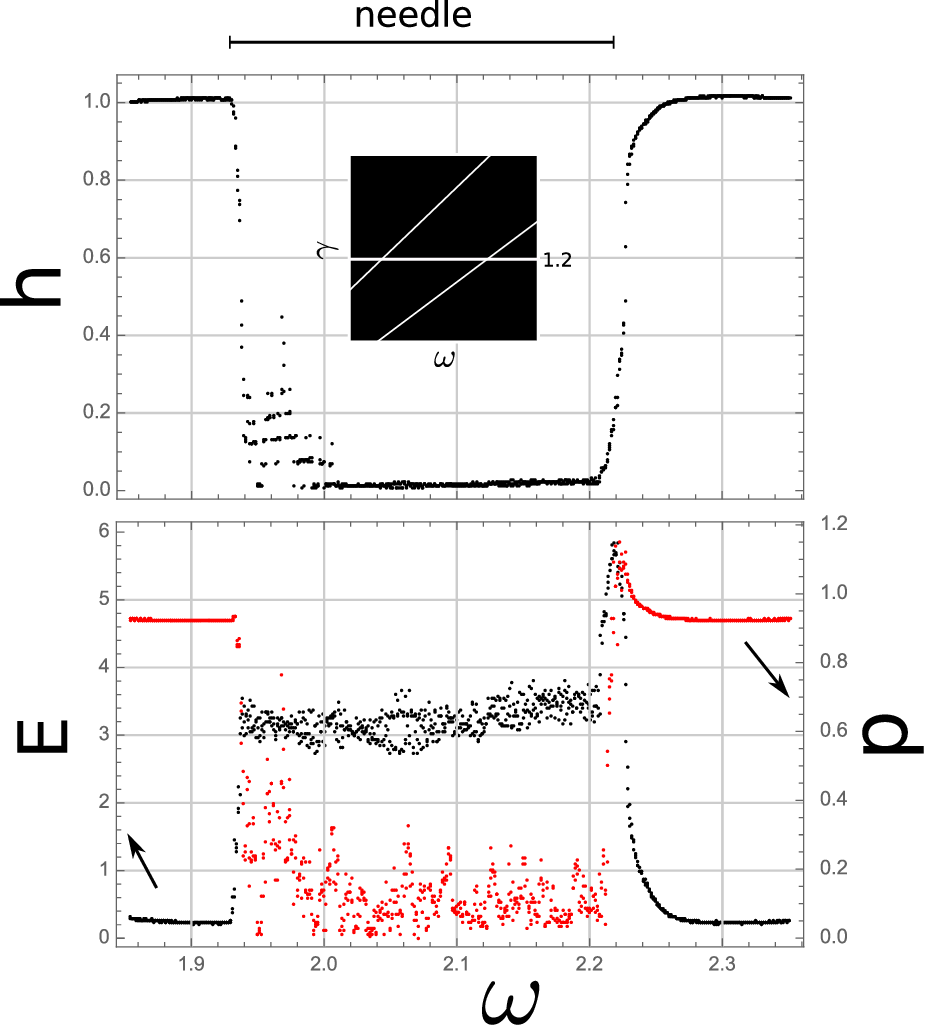}
\caption{Entropic measures along a line of constant $\gamma=1.2$ crossing the whole width of the needle region. The enhancement of computation at the right boundary can be seen as a surge in effective measure complexity (lower black plot) with a peak where the entropy density (upper plot) jumps towards maximum value.  $d$ (lower red plot) also has a jump with a peak, showing a transition to a chaotic regime. 
}\label{fig:needlew}
\end{figure}

In figure \ref{fig:needlew}, the entropic measures along a line of constant $\gamma=1.2$ are shown that goes through the whole width of the needle region. As $ \omega $ increases, the system goes from the chaotic region into the needle. A sudden drop of entropy density (upper plot of figure \ref{fig:needlew}) marks the transition, the effective measure complexity has a jump at the same transition, but no peak can be seen. The sensitivity to initial conditions, monitored by $d$, drops in the needle region to values below the chaotic regime but has an irregular behavior within the needle region for increasing values of $\omega$. A transition region in the left boundary is seen as irregular variations of $h$, consistent with the map shown in figure \ref{fig:entropicpd}. At the right boundary, when $h$ jumps to higher values, $E$ has a surge. This surge has been taken before as evidence of enhancement of computation at the edge of chaos \cite{estevez19}. This behavior is absent on the left boundary.

\begin{figure}[!t]
\centering
\includegraphics*[scale=0.5]{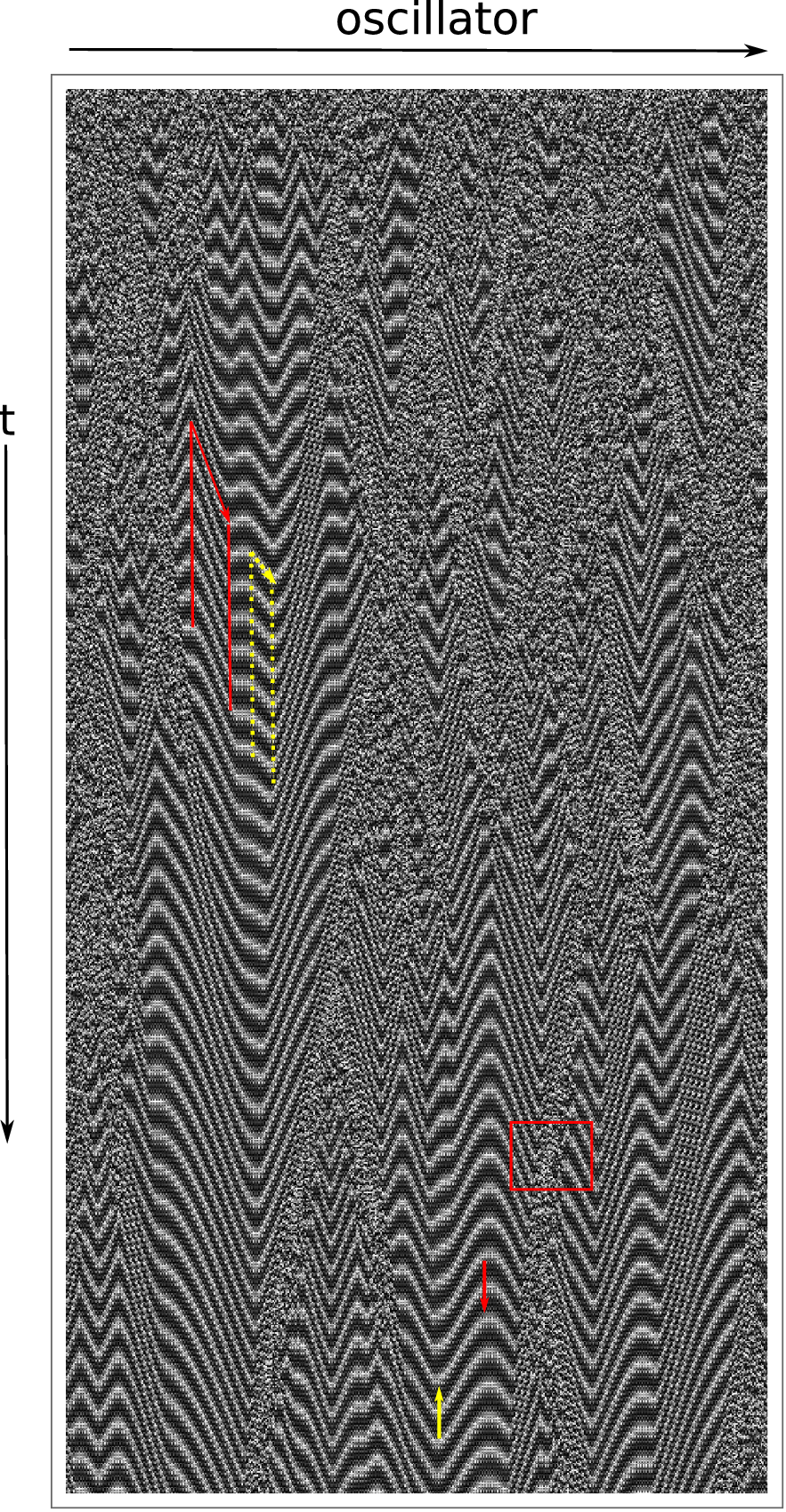}
\caption{Spatiotemporal map with random initial condition of a point at the lower boundary of the needle region at the onset of chaos ($\omega=2.225$, $\gamma=1.206$). The values of the plot have not been binarized. Gray levels correspond to values of the activity, $\sin \theta$, where black corresponds to $0$. Time is on the vertical scale, and spatial arrangement is along the horizontal scale. The red arrow points down to an emerging feature that propagates in both directions. The up, yellow arrow shows the collision of two features. The red rectangle shows a boundary region separating two wave regions where coherence is lost. On the left side, the diagonal arrows show two features traveling at different speeds and lasting different time-lapses. The left red arrow points to a slower translation speed, while the yellow one points to a faster translation speed, yet the first lasts longer than the latter. 
}\label{fig:detail}
\end{figure}

Figure \ref{fig:detail} is the spatiotemporal map of a point in the right boundary ($\omega=2.225$, $\gamma=1.206$) where $E$ has a peak. The map has not been binarized, so grey levels correspond to activity values $\sin \theta$, where the black color is for $0$. The map shows complex features with areas showing a wave-like solid pattern. The waves start at a local feature that emerges (straight down red arrow at the lower right points to one of such features) and spatially propagates in both directions with time. Two such propagating features can collide and annihilate (the straight, up, yellow arrow at the lower right points to one of such collisions). Features in the wave-like region can propagate at different speeds and last for different time-lapses, as can be seen in the upper left, red, and yellow diagonal arrows of the same figure \ref{fig:detail}. The boundaries between wavy regions seem incoherent. They can be both a source or a sink of propagating features that travel into or from the wave regions (right lower rectangle). The spatiotemporal map for this boundary region shows a width larger than one oscillator that propagates and splits with time. 

\section{Discussion}\label{sec:discus}

The left boundary of the needle region is close to a straight line parallel to the right boundary for $ \omega $ values larger than $2.33$. For lower values of $\omega$, the slope of the upper boundary is larger than that of the lower boundary. A straight line drawn for the middle of the needle region has a slope of $0.869$, which implies that as $ \omega $ increases,  the ratio $ \gamma/\omega $ decreases. For a decoupled Adler equation, the region $\gamma < \omega$ has no stable point, and $\theta$ is periodic with period $T=2\pi/\sqrt{\omega^2-\gamma^2}$. For such decoupled oscillator, for increasing $\omega$ values, $T\rightarrow 2\pi/\omega$, the self-feedback gets weaker, which can be readily seen looking at equation (\ref{eq:adler}) and ignoring the third term in the right side. 

The needle region lies, except for its apex, outside the equilibrium zone of the oscillator \cite{alonso17}, and locking of the phases is not possible; the phases of all oscillators continue to change with time. Within the whole needle region, structuring, as measured by the effective measure complexity, happens, pointing to correlations between the oscillators beyond the short range of the interaction. In the coupled oscillator system, as $\omega$ and $\gamma$ increase, the coupling gets weaker with regard to the self-feedback and the natural frequency. However,  in the studied region, the coupling is still a significant factor for the system. As a result, the periodic behavior of the isolated oscillators is perturbed by the feedback from the neighboring oscillators. These competing factors determine the behavior of each oscillator and the system. 
 
 The spatiotemporal diagrams within the needle region show wave-type patterns. Large coherent areas are found with vertical boundaries between them. Coherence is almost preserved for the whole range of oscillators which is verified by the dendrograms of figure \ref{fig:plot}b-1. The drop in entropy density as a function of time is a sign of irreversibility, and information is lost almost immediately from the initial condition. On the other hand, the increase of effective measure complexity measures the highly structured nature of the spatial configurations after the initial transient. Transmission of information across the whole arrangement of oscillators is a system feature. The almost periodic behavior of the system points to the storage of information in the time scale. The emergence of new features is limited to events that repeat in time. The difference with other points within the same region is the spatial wavelength and time-frequency, which govern the patterns' recurrence. For some $(\omega, \gamma)$ values, there is almost no evolution or a slow one for others. After the transient, short,  initial stage, the spatiotemporal map shows no high complexity and is far from chaotic.
 
 What is the difference with the critical region at the right boundary as the system moves towards the chaotic regime with increasing $\omega$? At the onset of chaos, the spatiotemporal map shows a wave-like pattern again. However, the wave patterns are now disrupted by the boundaries of several oscillators that seem not coherent with their neighbors. As a result, spatial coherence is lost (figure \ref{fig:detail}). The transmission of information across all oscillators seems to become unfeasible. In this critical region, local clusters of oscillators that behave coherently are found, which are isolated from other blocks of coherent oscillators by incoherent non-constant boundaries, a picture consistent with the corresponding dendrogram of figure \ref{fig:plot}b-2. The boundaries are seen to move as the system evolves, and new boundaries can emerge from existing ones. When a boundary splits in two, a new coherent region of wave pattern emerges. When two incoherent boundaries collide, a coherent wave region disappears. The $h$ plot vs. time shows that there is also a loss of information from the initial random configuration in this onset of chaos, but smaller than in the needle region. The spatiotemporal map further shows that not all initial information is lost as boundaries emerge from the initial configuration and persist in time.

On the other hand, the effective measure complexity is consistent with the qualitative picture; structuring occurs but not to the level of the needle region, and some randomness is kept in the configurations. In the evolution of the spatial configurations, there is a competing effect between memory preservation (given by $E$) and information production (given by $h$). All these features point to a complex map of competing patterns where information storage happens for a finite interval of time steps. The transmission of information occurs within blocks of oscillators and not through the whole arrangement and emergence of new regions, a characteristic feature of the system.

One may wonder if such rich behavior at the onset of chaos may allow for complex computations given the appropriate initial conditions.

\section{Conclusions}\label{sec:conclusions}

Coupled non-linear oscillators have been used extensively to model various complex systems, including brain functioning. It is common to assume that coupling goes beyond local links, and all-to-all topology and small-world topology have been studied. There is an idea that local couplings are too simple to give rise to complex behavior emerging from the collective correlation between a large enough number of units.  

Adler-type locally coupled oscillators was a toy model introduced by Alonso \cite{alonso17} that showed complex behavior in some areas of the phase diagram despite its very simple coupling. This should not have come as a surprise as elementary cellular automata already have given enough evidence of complex behavior emerging from nearest-neighbor couplings of simple units. 

Of the two regions with possible complex behavior, the so called needle region has been the focus of this contribution. It had already been reported that in the lower boundary of the needle, a jump of the different entropy magnitudes could signal enhanced computational capabilities at the onset of chaos. In this contribution it has been shown that within the needle region, a rich set of behaviors can be identified, and global coupling of the oscillator sets in after a transient period. This coupling is not the usually reported phase locking and can not be seen by simple distance measures such as the Hamming field. It is a coupling revealed as a correlation in the pattern-producing ability of one oscillator compared to the other. That is, two correlated oscillators mean that each produces similar sets of temporal patterns.  

A further finding is that the needle region is not homogeneous. The different spatiotemporal maps point out that the differences are in the spatial wavelength and time-frequency of the recurrence of wave-like patterns seen as almost coherent across the whole arrangement of oscillators. 

At the onset of chaos, the system exhibits local coupling seen in the spatiotemporal maps and verified by the distance dendrogram. This coexistence of local communities that are weakly coupled to other local communities is a hallmark of enhanced computational capabilities.  The intermediate values of entropy density and effective measure complexity for this region, a known feature for complex, capable information processing systems, further emphasize this improved capability.

To conclude with a comment, the fact that such a simple system can give rise to such rich sets of behavior could point to the idea that one could expect even more richness in more realistic sophisticated models.

\section{Acknowledgments}

The MPI-PKS partially financed this work under their visitor program scheme. Partial support came from the University of Havana.


\end{document}